# Terahertz Radiation Detection by Field Effect Transistor in Magnetic Field

S. Boubanga-Tombet[1], M. Sakowicz[1,3], D. Coquillat[1], F. Teppe[1], W. Knap[1,3]

[1]Université Montpellier II and CNRS – GES –UMR5650 Place E.Bataillon 34095 Montpellier, France

M. I. Dyakonov[2]

[2]Laboratoire de Physique Théorique et Astroparticules, Université Montpellier II, CNRS, Place E.Bataillon 34095 Montpellier, France

K. Karpierz[3], J. Łusakowski[3], M.Grynberg[3]

[3]Institute of Experimental Physics, University of Warsaw, Hoża 69, 00-681 Warsaw, Poland

We report on terahertz radiation detection with InGaAs/InAlAs field effect transistors in quantizing magnetic field. The photovoltaic detection signal was investigated at 4.2 K as a function of the gate voltage and magnetic field. Oscillations analogous to the Shubnikov-de Haas oscillations as well as their strong enhancement at the cyclotron resonance were observed. The results are quantitatively described by a recent theory, showing that the detection is due to rectification of the terahertz radiation by plasma waves related nonlinearities in the gated part of the channel.

Field effect transistors have been shown to be efficient detectors of electromagnetic radiation of THz and sub-THz frequencies operating at liquid helium and at room temperatures [1-4]. Resonant and nonresonant detection of THz radiation by high electron mobility transistors (HEMTs) was demonstrated in commercial GaAs/AlGaAs [5-9] and InGaAs based HEMTs [10-12]. These HEMTs were found to have competitive parameters in comparison with other kinds of available detectors. Detection of THz radiation by Si-CMOS based transistors was also demonstrated and already used in many application systems such as terahertz imaging and cameras [13-15].

The detection signal results from the rectification of THz currents induced by the incident radiation in the transistor channel. The rectification takes place due to a nonlinear response of the gated two dimensional electron gas. This is due to the superposition of two effects: i) radiation induced modulation of the carrier density in the channel and ii) modulation of electron drift velocity. The photoresponse is enhanced close to the plasma resonances.

In a high magnetic field the conduction band splits into the Landau levels, which results in Shubnikov-de Haas (SdH) oscillations and the cyclotron resonance (CR). This can influence the photovoltaic response. Recently, oscillations of the detection signal in magnetic field as well as the strong increase of the signal in the vicinity of the cyclotron resonace have been reported [16].

Although both CR and SdH effects were experimentally observed, it was unclear what the exact physical mechanism is and which part of the transistor is responsible for the observed effects. The difficulty of the interpretation was due to the fact that the influence of the gate voltage on the observed signal could not be clearly established, because the investigated transistors had a long ungated region (~9.2 µm) and a short gated region (~0.8 µm). Hence the gate controls the electron density only in a small part of the channel.

The purpose of this work is to clarify the physical mechanism of SdH and CR related detection by FETs. We investigate the THz response in the transistor of a simple architecture, with the gate covering the major part of the channel. In this simplified structure, the effect of the gate bias on the observed signal can be clearly established and the results can be compared with the recent theory [17].

Experiments were carried out at 4.2 K on a pseudomorphic InGaAs/InAlAs HEMT grown and processed in IEMN, Lille. Details of heterostructure and HEMT fabrication are described in Ref. 12. Room temperature mobility and sheet carrier density were 11500 cm²/V·s and $2.5 \cdot 10^{12}$ cm$^{-2}$, respectively. The gate length, gate width, and channel length were 1.5 µm, 25 µm, and 2.6 µm, respectively. Thus the gate covered about 60% of the channel and the transistor geometry was close to that theoretically considered in Ref. 17.

The radiation of frequency 2.5 THz was generated by an optically pumped molecular CH$_3$OH laser. The laser beam was modulated by a mechanical chopper (137 Hz) and guided to the sample through inox light pipes. The beam was neither intentionally focalized nor polarized. The transistor was placed at the center of a superconducting coil in the Faraday configuration and cooled by a helium exchange gas. The photovoltaic signal, $\Delta U$, was measured between source and drain contacts using a standard lock-in technique.

Two types of experiments were performed. First, the signal was measured as a function of the gate voltage $V_g$ at fixed magnetic field $B$. Second, the magnetic field was swept at a fixed gate voltage. The results are shown in Figs. 1 - 3. In both kinds of experiments the photoresponse shows an oscillatory behavior. Its periodicity versus $V_g$ and $1/B$, clearly indicates that the oscillations are related to the crossing between the Fermi level and the Landau levels (analogous to the SdH effect). Additionally, a strong enhancement of the signal is observed close to $B_c = 4.5$ T (CR condition).

The interpretation of the experimental data is based on the recently developed model of THz detection by transistors with the gated plasma in quantizing magnetic field [17]. The theory provides an analytical result describing the influence of CR and SdH effects on the photoresponse:

$$\Delta U = \frac{1}{4}\frac{U_a^2}{U_o}\left[f(\beta) - \frac{d\gamma}{dn}\frac{n}{\gamma}g(\beta)\right], \quad (1)$$

where $n$ is the electron concentration, $U_a$ is ac gate-to-source voltage induced by the radiation, $U_0$ is the swing voltage, $\beta=\omega_c/\omega$ is the magnetic field in units its resonant value, $B_c$, for a given radiation frequency $\omega$. and $\omega_c$ is the cyclotron frequency.

The parameter $\gamma$ is an oscillatory function of the electron concentration (or gate voltage) and the magnetic field:

$$\gamma = \gamma_0\left[1 - 4\frac{\chi}{\sinh\chi}\exp\left(-\frac{\pi}{\omega\tau_q}\frac{1}{\beta}\right)\cos\left(-\frac{2\pi}{\beta}\frac{E_F}{\hbar\omega}\right)\right], \quad (2)$$

where $\chi = 2\pi^2 kT/\hbar\omega_c$, $\tau_q$ is the quantum relaxation time, defining the damping of the SdH oscillations, and $E_F$ is the Fermi energy, which is proportional to the electron concentration.

The function $f$ represents the non-oscillating part of the detection signal resulting from the previously analyzed sources of nonlinearity [2]. The second term, proportional to $d\gamma/dn$, comes from additional nonlinearities related to the oscillatory dependence of the mobility on the electron concentration. The envelope for the oscillating part of the signal is described by $g(\beta)$. The $f$ and $g$ functions are given by:

$$f(\beta) = 1 + \frac{1+F}{\sqrt{\alpha^2+F^2}}, \quad (3)$$

$$g(\beta) = \frac{1+F}{2}\left(1 + \frac{F}{\sqrt{\alpha^2+F^2}}\right). \quad (4)$$

Here $F$ depends only on $\beta = \omega_c/\omega$ and $\alpha = (\omega\tau)^{-1}$, with $\tau$ being the momentum relaxation time:

$$F = \frac{1+\alpha^2-\beta^2}{1+\alpha^2+\beta^2}, \quad (5)$$

Fig. 1$a$ shows a waterfall plot of the detection signal as a function of the gate voltage at different values of $B$. As $B$ increases, the amplitude of oscillations grows (while $B<B_c$). This can be understood as a result of increasing nonlinearity related to SdH oscillations described by $d\gamma/dn$ term in Eq. 1. In the post-cyclotron resonance region ($B>B_c$), the oscillations of the signal are damped. This can be qualitatively explained by considering the plasma wave dispersion law in magnetic field:

$$k = \frac{\omega}{s}\sqrt{1 - \frac{\omega_c^2}{\omega^2}}, \quad (6)$$

where $s=(eU_0/m)^{1/2}$ is the plasma wave velocity at $B=0$. For $\omega_c > \omega$ the plasma wave vector becomes imaginary, so that plasma waves cannot propagate, and one can expect a reduction of the signal, as seen in Fig. 1$a$.

Fig. 1$b$ shows calculation of the photoresponse as a function of $N = E_F/\hbar\omega = \pi\hbar n/eB_c$, which is the number of Landau levels below the Fermi level at cyclotron resonance. We took the following values: $\chi = 0.7$ (corresponding to $T = 4.2$ K, and $\omega = 2\pi \cdot 2.5$ THz), $\omega\tau_q = 0.37$, and $\omega\tau = 5.6$

(corresponding to the mobility of 15000 cm$^2$/(V·s) at 4.2 K, estimated from independent magnetoresistance measurements).

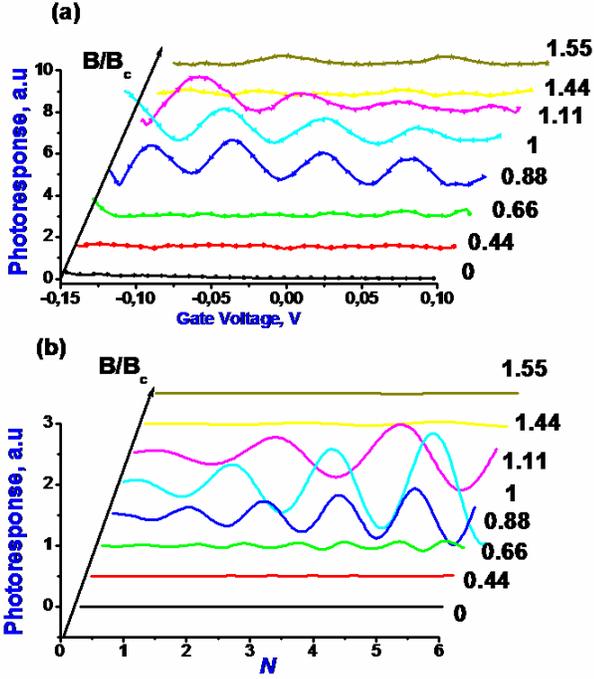

FIG. 1. Waterfall plot of the photoresponse at 2.5 THz as a function of the gate voltage for different values of the magnetic field at T = 4.2 K. Values of $B/B_c$ are indicated. *a* - experimental,
*b* - calculations using Eq. 1, the parameter *N* is proportional to the gate voltage.

The experimental results are reproduced quite well by the theoretical curves. The strong dependence of the photoresponse on the gate voltage indicates that the rectification takes place in the gated part of the transistor channel.

A more detailed analysis of the data is presented in Fig. 2. In fact, in Fig 1*a*, not only the oscillation amplitude, but also the constant background of the signal, obtained by averaging over oscillations, changes with *B*. This can be seen more clearly in Fig. 2 which shows the signal averaged over oscillations as a function of the magnetic field (stars). The maximum is close to the CR condition. This averaged signal dependence is well reproduced by the theory [17], using Eq. 4 with the same parameters as above (see Fig. 2, bottom).

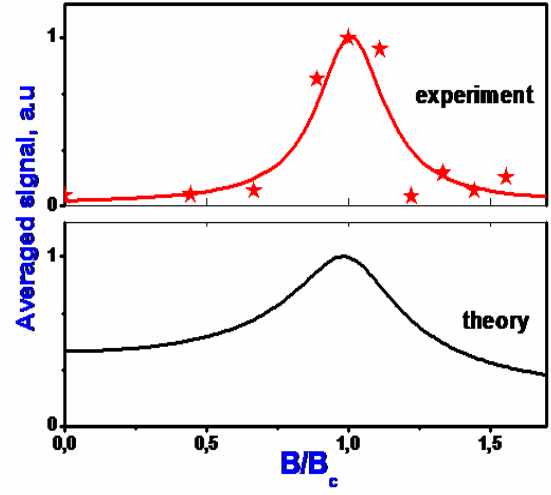

FIG. 2: Top: photoresponse averaged over oscillations (stars) extracted from Fig. 1 data. Bottom: calculations using Eqs. (1) and (4).

Fig. 3*a* shows the measured magnetic field dependence for fixed gate voltage (fixed electron density). Fig. 3*a* corresponds to a relatively high density, for which one can see strong oscillatory behavior of the photovoltaic signal. Fig. 3*b* corresponds to a lower density with a very week oscillations signature. For $B>B_c$ the oscillation amplitude decreases. This behavior is once again described by the theory [17], where the envelope of the oscillating part, $g(\beta)$, exhibits a fast decay beyond the cyclotron resonance, in accordance with Eq. 6. The solid line in Fig. 3*a* (bottom) shows calculation of the photoresponse as a function of the reduced magnetic field $\beta$ for $\omega\tau = 5.6$, $\omega\tau_q = 0.37$. The carrier density was taken to be $n = 1.141\cdot10^{12}$ cm$^{-2}$ as obtained from magneto-resistance measurements. This corresponds to *N*=5.25. A good agreement between the experimental data and the calculations can be seen. It can be also seen in Fig. 3*b* where the bottom curve shows calculations for $\omega\tau = 5.6$, $\omega\tau_q = 0.5$ and $N = 2.3$. A strong enhancement around cyclotron resonance is seen. This effect can be used for magnetically tunable terahertz detection.

In conclusion, we have performed magnetic field studies of plasma wave related detection in InGaAs/InAlAs field effect transistors. We observe oscillatory behavior of the photovoltaic signal similar to the Shubnikov-de Haas effect, as well as a strong enhancement of the signal at cyclotron resonance. The experimental data are quantitatively reproduced by the theory [17]. Our results clearly show that the detection mechanism is related to the THz rectification by nonlinearities of the plasma in the gated region of the transistor.

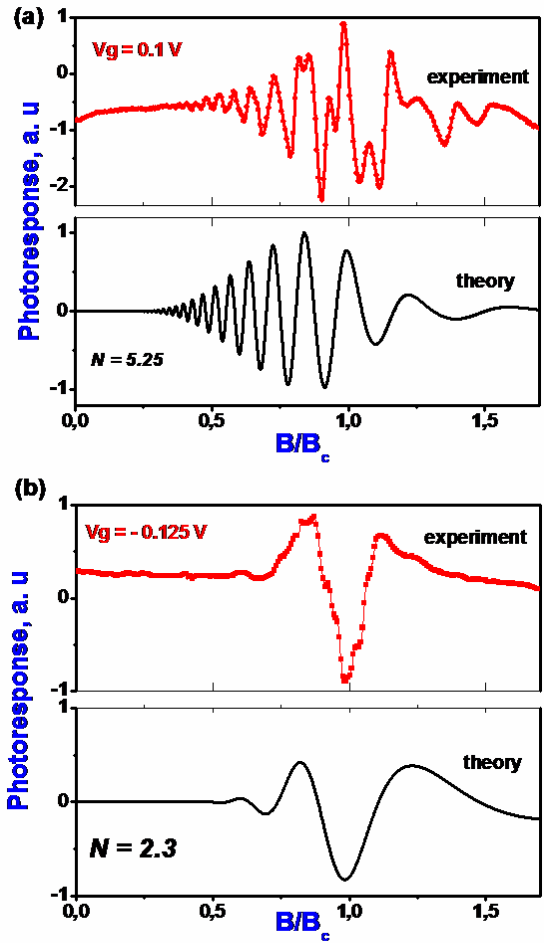

FIG. 3. *a* - Top: experimental photoresponse as a function of the magnetic field for higher values of the electron density ($V_g$ = 0.1 V) at 4.2 K. Bottom: calculations using Eq. (1). *b* - Same as in *a* for lower carrier density ($V_g$ = -0.125 V)

*Acknowledgements:* The transistors used in this work was produced by the IEMN of Lille. We acknowledge the help of A. Cappy, S. Bollaert, A. Shchepetov. and M. Lifshits. This work was supported by CNRS, the GDR-E project "Semiconductor sources and detectors of terahertz frequencies", and by the joint French-Lithuanian research program "Gilibert/EGIDE." We acknowledge the region of Languedoc-Roussillon through the "Terahertz Platform" project and the European Union Grant No. MTKD-CT-2005-029671.